\documentclass[a4paper]{article}

\usepackage{amsmath,amssymb}
\usepackage{INTERSPEECH2020}
\usepackage{multirow}
\usepackage{color}
\usepackage{subcaption}
\usepackage{hyperref}

\DeclareMathOperator{\cossim}{sim}
\DeclareMathOperator{\score}{score}

\usepackage{xcolor}

\definecolor{pink}{RGB}{219, 48, 122}

\usepackage{bm}
\usepackage{amssymb,amsmath,graphicx,bbm,color,booktabs}
\usepackage{amsopn}

\newcommand{\vu}{\bm{u}}               
\newcommand{\vv}{\bm{v}}

\newcommand{\R}{\mathbb{R}}

\newcommand{\X}{\mathcal{X}}
\newcommand{\Z}{\mathcal{Z}}

\newcommand{\z}{\mathbf{z}}

\renewcommand{\eqref}[1]{Eq.~(\ref{#1})}

\def \x{{\mathbf x}}

\title{Self-Supervised Contrastive Learning for Unsupervised Phoneme Segmentation}
\name{Felix Kreuk$^1$, Joseph Keshet$^1$, Yossi Adi$^2$}

\address{
  $^1$Bar-Ilan University\\
  $^2$Facebook AI Research}
\email{felix.kreuk@gmail.com}

\begin{document}

\maketitle
\begin{abstract}
We propose a self-supervised representation learning model for the task of unsupervised phoneme boundary detection. The model is a convolutional neural network that operates directly on the raw waveform. It is optimized to identify spectral changes in the signal using the Noise-Contrastive Estimation principle. At test time, a peak detection algorithm is applied over the model outputs to produce the final boundaries. As such, the proposed model is trained in a fully unsupervised manner with no manual annotations in the form of target boundaries nor phonetic transcriptions. We compare the proposed approach to several unsupervised baselines using both TIMIT and Buckeye corpora. Results suggest that our approach surpasses the baseline models and reaches state-of-the-art performance on both data sets. Furthermore, we experimented with expanding the training set with additional examples from the Librispeech corpus.
We evaluated the resulting model on distributions and languages that were not seen during the training phase (English, Hebrew and German) and showed that utilizing additional untranscribed data is beneficial for model performance.
Our implementation is available at: \textcolor{pink}{\url{https://github.com/felixkreuk/UnsupSeg}}.\\

\end{abstract}

\section{Introduction}
\vspace{0.2cm}
\textit{Phoneme Segmentation} or \textit{Phoneme Boundary Detection} is an important precursor task for many speech and audio applications such as Automatic Speech Recognition (ASR) ~\cite{kubala1996transcribing, rybach2009audio, yeh2018unsupervised}, speaker diarization~\cite{moattar2012review}, keyword spotting~\cite{keshet2009discriminative}, and speech science~\cite{adi2016vowel,adi2015vowel}. 

The task of phoneme boundary detection has been explored under both supervised and unsupervised settings \cite{kreuk2020phoneme, franke2016phoneme, michel2016blind, rasanen2014basic}. Under the supervised setting two schemes have been considered: \textit{text-independent} speech segmentation and \textit{phoneme-to-speech alignment} also known as \emph{forced alignment}, which is a text-dependent task. In the former setup, the model is provided with target boundaries, while in the latter setup, the model is provided with additional information in the form of a set of pronounced or presumed phonemes. In both schemes, the goal is to learn a function that maps the speech utterance to the target boundaries as accurately as possible.

However, creating annotated data of phoneme boundaries is a strenuous process, often requiring domain expertise, especially in low-resource languages~\cite{goldrick2016automatic}. As a consequence, unsupervised methods and Self-Supervised Learning (SSL) methods, in particular, are highly desirable and even essential. 

In unsupervised phoneme boundary detection, also called \textit{blind-segmentation}~\cite{rasanen2011blind, michel2016blind}, the model is trained to find phoneme boundaries using the audio signal only. In the self-supervised setting, the unlabeled input is used to define an auxiliary task that can generate labeled pseudo training data. This can then be used to train the model using supervised techniques. SSL has been proven to be effective in natural language processing~\cite{devlin2018bert, liu2019roberta}, vision~\cite{chen2020simple}, and recently has been shown to generate a useful representation for speech processing~\cite{oord2018representation, schneider2019wav2vec}.

\begin{figure}[t]
  \centering
  \includegraphics[width=\linewidth]{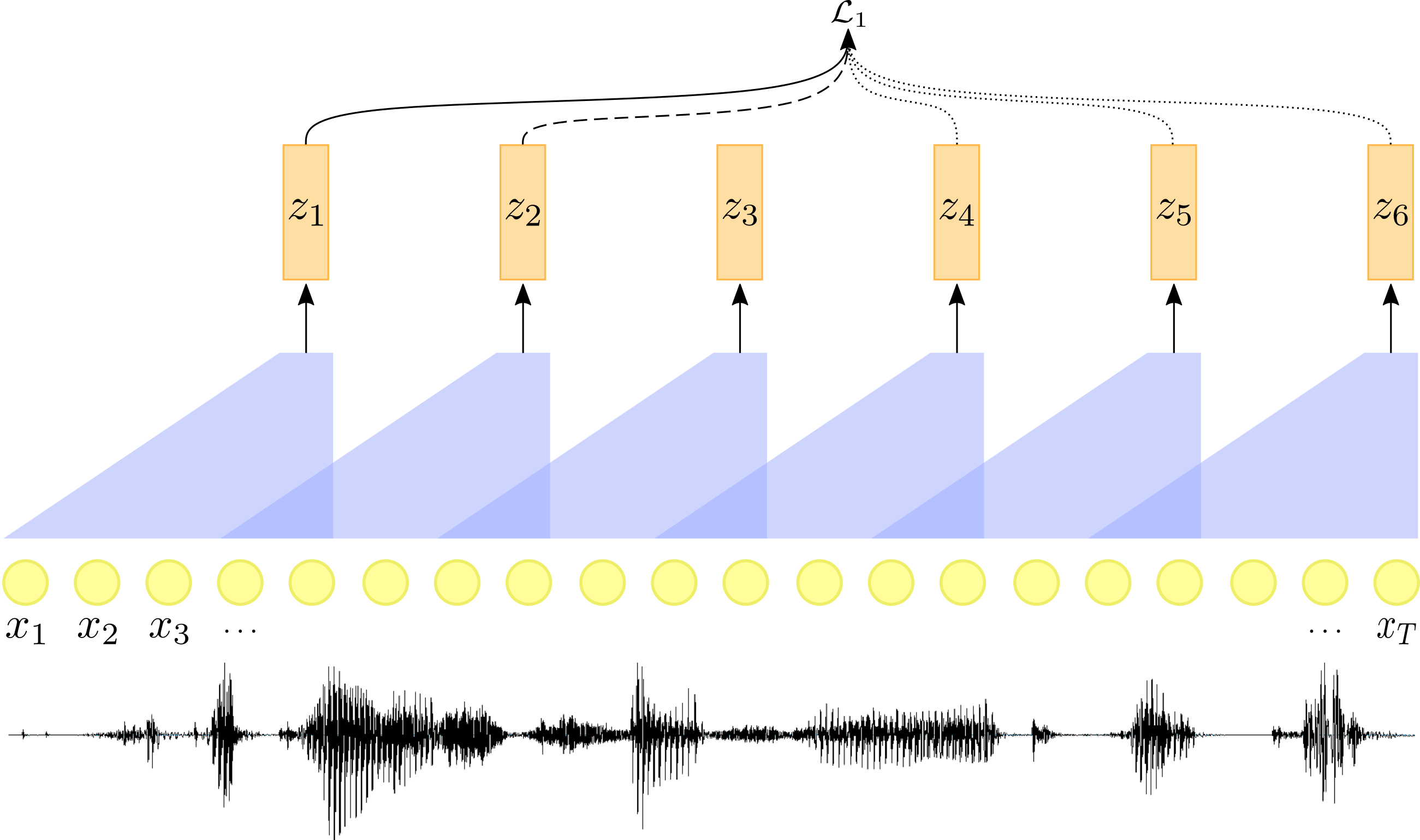}
  \caption{An illustration of our model and SSL training scheme. The solid line represents a reference frame $\z_1$, the dashed line represents its positive pair $\z_2$, and the dotted lines represent negative distractor frames randomly sampled from the signal.}
  \label{fig:model}
\end{figure}

Most of the SSL work in the domain of speech processing and recognition has been focused on extracting acoustic representations for the task of ASR \cite{oord2018representation, schneider2019wav2vec}. However, it remains unclear how effective SSL methods are when applied to other speech processing applications.

In this work, we explore the use of SSL for phoneme boundary detection. Specifically, we suggest learning a feature representation from the raw waveform to identify spectral changes and detect phoneme boundaries accurately. We optimize a Convolutional Neural Network (CNN) using the Noise Contrastive Estimation principle~\cite{gutmann2010noise} to distinguish between pairs of \textit{adjacent frames} and pairs of random distractor pairs. The proposed model is depicted in Figure~\ref{fig:model}. During inference, a peak-detection algorithm is applied over the model outputs to produce the final segment boundaries.

We evaluate our method on the TIMIT~\cite{garofolo1993timit} and Buckeye~\cite{pitt2005buckeye} datasets. Results suggest that the proposed approach is more accurate than other state-of-the-art unsupervised segmentation methods. We conducted further experiments with larger amount of untranscribed data that was taken from the Librispeech corpus. Such an approach proved to be beneficial for better overall performance on unseen languages.
\paragraph*{Our contributions:}
\begin{itemize}
    \item We demonstrated the efficiency of SSL, in terms of model performance, for learning effective representations for unsupervised phoneme boundary detection.
    \item We provide SOTA results in the task of unsupervised phoneme segmentation on several datasets.
    \item We provide empirical evidence that leveraging more unlabeled data leads to better overall performance on unseen languages.
\end{itemize}

\noindent The paper is organized as follows: In Section~\ref{sec:model} we formally set the notation and definitions used throughout the paper as well as the proposed model. Section~\ref{sec:exp} provides empirical results and analysis. In Section~\ref{sec:rel} we refer to the relevant prior work. We conclude the paper with a discussion in Section~\ref{sec:dis}.
\section{Related work}
\label{sec:rel}

The task of phoneme boundary detection was explored in various settings. Under the supervised setting, the most common approach is the forced alignment. In this setup, previous work mainly involved hidden Markov models (HMMs) or structured prediction algorithms on handcrafted input features~\cite{keshet2005phoneme, mcauliffe2017montreal}. In the text independent setup, most previous work reduced the task of phoneme segmentation to a binary classification at each time-step~\cite{king2013accurate, franke2016phoneme}. More recently, \cite{kreuk2020phoneme} suggested using an RNN-coupled with structured loss parameters.

Under the unsupervised setting, the speech utterance is provided by itself with no boundaries as supervision. Traditionally, signal processing methods were used to detect spectral changes over time~\cite{dusan2006relation, estevan2007finding, almpanidis2008phonemic, rasanen2011blind}, such areas of change were presumed to be the boundary of a speech unit. Recently, Michel \emph{et al.}~\cite{michel2016blind} suggested training a next-frame prediction model using HMM or RNN. Regions of high prediction error were identified using peak detection and flagged as phoneme boundaries. More recently, Wang \emph{et al.}~\cite{wang2017gate} suggested training an RNN autoencoder and tracking the norm of various intermediate gate values (forget-gate for LSTM and update-gate for GRU). To find phoneme boundaries, similar peak detection techniques were used on the gate norm over time.

In the field of self-supervised learning, Van Den Oord \emph{et al.} \cite{oord2018representation} and Schneider \emph{et al.} \cite{schneider2019wav2vec} suggested to train a Convolutional neural network to distinguish true future samples from random distractor samples using a probabilistic contrastive loss. Also called Noise Contrastive Estimation, this approach exploits unlabeled data to learn a representation in an unsupervised manner.
The resulting representation proved to be useful for a variety of downstream supervised speech tasks such as ASR and speaker identification.

\section{Model}
\label{sec:model}

Following the recent success of contrastive self-supervised learning \cite{chen2020simple, oord2018representation, schneider2019wav2vec}, we propose a training scheme for learning useful representations for unsupervised phoneme boundary detection. We denote the domain of audio samples by $\X \subset \R$. The representation for a raw speech signal is therefore a sequence of samples $\x = (x_1,\ldots, x_T)$, where  $x_t\in\X$ for all $1\leq t \leq T$. The length of the input signal varies for different inputs, thus the number of input samples in the sequence, $T$, is not fixed. We denote by $\X^*$ the set of all finite-length sequences over $\X$.

Denote by $\z = (\z_1,\ldots,\z_{L})$ a sequence of spectral representations sampled at a low frequency. Each element in the sequence is an $N$-dimensional real vector, $\z_i\in \Z\subseteq\R^N$ for $1\le i \le L$. Every element $\z_i$ corresponds to a 10 ms frame of audio with a processing window of 30 ms. Let $\Z^*$ denote all finite-length sequences over $\Z$.

We learn an encoding function $f: \X^* \to \Z^*$, from the domain of audio sequences to the domain of spectral representations. The function $f$ is optimized to distinguish between pairs of \textit{adjacent} frames in the sequence $\z$ and pairs of randomly sampled distractor frames from $\z$.
Denote by $D(\z_i)$ the set non-adjacent frames to $\z_i$, 
\begin{equation}
\label{eq:negative_set}
    D(\z_i) = \{\z_j : |i-j| > 1 ~,~ \z_j \in \z \}. 
\end{equation}
Practically we use $K$ randomly selected frames from $D(\z_i)$, and denote it by $D_K(\z_i)\subset D(\z_i)$. The loss for frame $\z_i$ is defined as,
\begin{equation}
    \hat{\mathcal{L}}(\z_i, D_K(\z_i)) = -\log \frac{{e}^{\cossim(\z_i,\z_{i+1})}}{\sum_{\z_j \in \{\z_{i+1}\} \cup D_K(\z_i)} {e}^{\cossim(\z_i,\z_j)}},
\end{equation}
where $\cossim(\vu,\vv)=\vu^\top \vv /||\vu||\,||\vv||$ denotes the cosine similarly between two vectors $\vu$ and $\vv$. 
Overall, given a training set of $m$ examples $S = \{\x_i\}_{i=1}^m$, we would like to minimize the following objective function,
\begin{align}
    \mathcal{L} = \sum_{\x \in S} \sum_{\z_i \in f(\x)} \hat{\mathcal{L}}(\z_i, D_K(\z_i))
\end{align}

During inference, we receive a new utterance $\x$. We then apply the encoding function to get $\z=f(\x)$. We set the score for a boundary at time $i$ to be the dissimilarity between the $i$-th frame and the $i+1$-th frame for $i = 1, \ldots, L-1$. That is 
\begin{equation}
    \label{eq:dis}
    \score({\z}_i) = -\cossim(\z_i, \z_{i+1})~.
    \end{equation}
Intuitively, $\score({\z}_i)$ can be interpreted as the model's confidence that the next frame $\z_{i+1}$ belongs to a different segment than that of the current frame $\z_i$. 
Thus, times with high dissimilarity values are associated with segment changes, and are considered as candidates for segment boundaries. 
We apply a peak detection algorithm over the dissimilarity values, $\score({\z})$ to get the final segmentation. The frames for which the score exceeds a peak prominence of $\delta$ are predicted as boundaries. The optimal value of $\delta$ is tuned in a cross-validation procedure.

\begin{figure}[t]
  \centering
  \includegraphics[width=\linewidth]{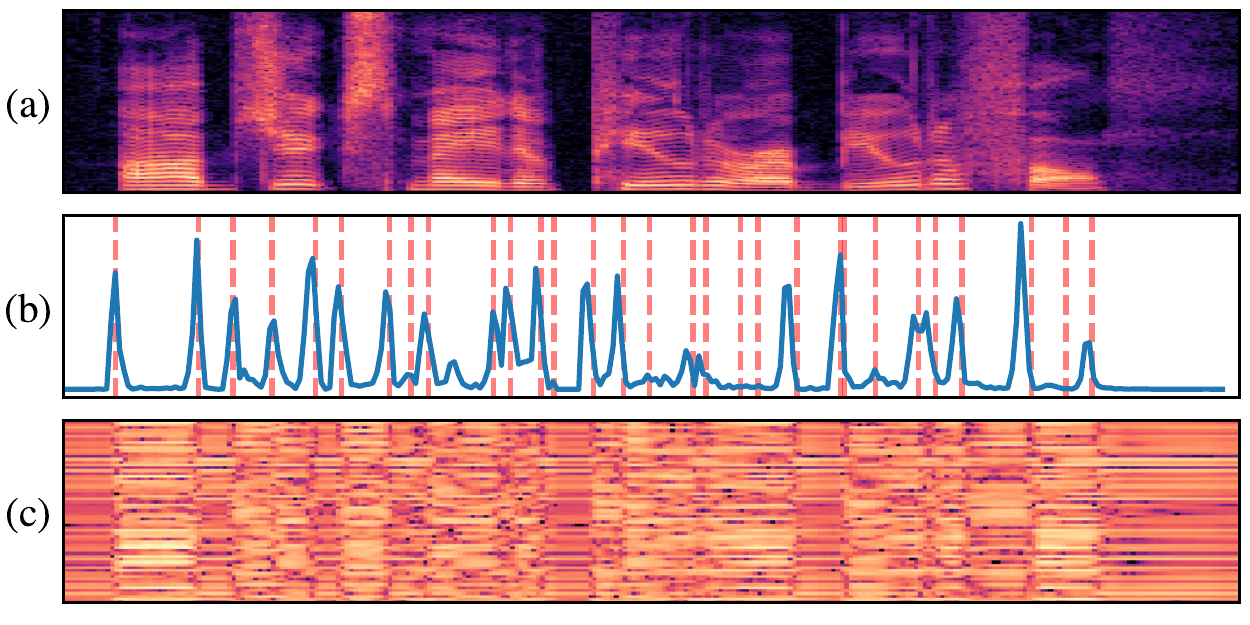}
  \caption{An illustration of the prediction produced by our model: (a) the original spectrogram; (b) our model's output at each time step, red dashed lines represent the ground truth segmentation; (c) the learned representation $\z$.}
  \label{fig:speech_production}
\end{figure}

Figure~\ref{fig:speech_production} presents an example utterance from TIMIT. The power spectrum of the utterance is presented in (a), the score function is presented in (b) and the corresponding learned representation $\z$ in (c).

\label{sec:exp}

\begin{table*}[t!]
    \renewcommand{\arraystretch}{1.1}
	\centering
	\caption{
    	Comparison of phoneme segmentation models using TIMIT and Buckeye data sets. Precision and recall are calculated with tolerance value of 20 ms. Results marked with * are reported using our own optimization.
    	}
  \vspace{5pt}
  \label{tab:phoneme_seg_results}
  \begin{tabular}{l|l|cccc|cccc}
	\toprule
	\multicolumn{2}{c}{} & \multicolumn{4}{c}{TIMIT} & \multicolumn{4}{c}{Buckeye}\\
	\midrule
	Setting & Model & Precision & Recall & F1 & R-val & Precision & Recall & F1 & R-val  \\
    \midrule
    \multirow{4}{*}{Unsupervised~~~}
    & Hoang \emph{et al.}~\cite{hoang2015blind}  & - & - &  78.20 & 81.10  & - & - & - & -  \\
	& Michel \emph{et al.}~\cite{michel2016blind} &  74.80 & 81.90 & 78.20 & 80.10 & 69.34$^*$ & 65.14$^*$ & 67.18$^*$ & 72.13$^*$  \\
    & Wang \emph{et al.}~\cite{wang2017gate}  & - & - &  - & 83.16  & 69.61$^*$ & 72.55$^*$ & 71.03$^*$ & 74.83$^*$  \\
    & \textbf{Ours}  & \textbf{83.89} &	\textbf{83.55} & \textbf{83.71} &	\textbf{86.02} & \textbf{75.78} & \textbf{76.86} & \textbf{76.31} &	\textbf{79.69} \\
    \midrule
    \multirow{3}{*}{Supervised}
    & King \emph{et al.}\cite{king2013accurate}  &  87.00 & 84.80 & 85.90 & 87.80 & - & - & - & -  \\
    & Franke \emph{et al.}\cite{franke2016phoneme}  & 91.10 & 88.10 &  89.6 & 90.80 & 87.80 & 83.30 &  85.50 & 87.17 \\
    & Kreuk \emph{et al.}\cite{kreuk2020phoneme}   & 94.03 & 90.46  & 92.22   & 92.79 & 85.40 & 89.12  & 87.23  & 88.76 \\
    \bottomrule
  \end{tabular}
\end{table*}

\section{Experiments}

In this section, we provide a detailed description of the experiments. We start by presenting the experimental setup. Then we outline the evaluation method. We conclude this section with experimental results and analysis.

\subsection{Experimental setup}

The function $f$ was implemented as a convolutional neural network, constructed of 5 blocks of 1-D strided convolution, followed by Batch-Normalization and a Leaky ReLU~\cite{maas2013rectifier} nonlinear activation function. The network $f$ has kernel sizes of $(10, 8, 4, 4, 4)$, strides of $(5, 4, 2, 2, 2)$ and 256 channels per layer. Finally, the output was linearly projected by a fully connected-layer. Overall the model was similar to the one proposed by~\cite{oord2018representation, schneider2019wav2vec}. However, unlike the aforementioned prior work, the proposed model does not utilize a \textit{context network}. Our experiments with such a network led to inferior performance, and therefore this component was omitted from the final model architecture. 

We optimized the model using a batch size of 8 examples and a learning-rate of 1e-4 for 50 epochs. We follow an early-stopping criterion computed over the validation set. All reported results are averaged over a set of 3 runs using cross-validation with different random seed values. To get $D_K$ we experimented $K \in \{1,3,5,7,10\}$, but did not observe significant differences in performance.

We evaluated our model on both TIMIT and Buckeye corpora. For the TIMIT corpus, we used the standard train/test split, where we randomly sampled ~10\% of the training set for validation. For Buckeye, we split the corpus at the speaker level into training, validation, and test sets with a ratio of 80/10/10. Similarly to~\cite{kreuk2020phoneme}, we split long sequences into smaller ones by cutting during noises, silences, and un-transcribed segments. Overall, each sequence started and ended with a maximum of 20 ms of non-speech\footnote{All experiments were conducted at Bar-Ilan university.}.

\subsection{Evaluation method}
Following previous work on phoneme boundary detection~\cite{michel2016blind, wang2017gate}, we evaluated the performance of the proposed models and baseline models using precision ($P$), recall ($R$) and F1-score with a tolerance level of 20 ms.

A drawback of the F1-score for boundary detection is its sensitivity to over-segmentation. A naive segmentation model that outputs a boundary every 40 ms may yield a high F1-score by achieving high recall at the cost of low precision. The authors in~\cite{rasanen2009improved} proposed a more robust complementary metric denoted as \emph{R-value}:
\begin{equation}
\begin{split}
&\textrm{R-value} = 1 - \frac{|r_1|+|r_2|}{2} \\
&r_1 = \sqrt{(1-R)^2 + (OS)^2}, ~~  
r_2 = \frac{-OS+R-1}{\sqrt{2}}
\end{split}
\end{equation}
where $OS$ is an over-segmentation measure, defined as $OS=R/P-1$. Overall the performance is presented in terms of Precision, Recall, F1-score and R-value.

\subsection{Results}
In Table~\ref{tab:phoneme_seg_results} we compared the proposed model against several unsupervised phoneme segmentation baselines: Hoang \emph{et al.}~\cite{hoang2015blind}, Michel \emph{et al.}~\cite{michel2016blind}, and Wang \emph{et al.}~\cite{wang2017gate}. We also report results for SOTA supervised algorithms in order to gauge the gap between the unsupervised and supervised methods.
As the unsupervised baselines did not report results for the Buckeye data set, and there are no pre-trained models available, we optimized these models locally. 
For a fair comparison we verified that the performance of the reproduced models is comparable to the one originally reported on TIMIT. These results are marked with *.

Results suggest that the proposed model is superior to the baseline models over all metrics on both corpora. 
Notice, for the TIMIT benchmark, the proposed model achieves comparable results to a supervised method based on a Kernel-SVM~\cite{king2013accurate}.
Additionally, as opposed to the reported unsupervised baselines which are built using Recurrent Neural Networks, our model is mainly composed of convolutional operations, hence can be parallelized over the temporal axis.

\begin{figure*}[t!]
     \centering
     \begin{subfigure}[b]{0.24\linewidth}
         \centering
         \includegraphics[width=\columnwidth]{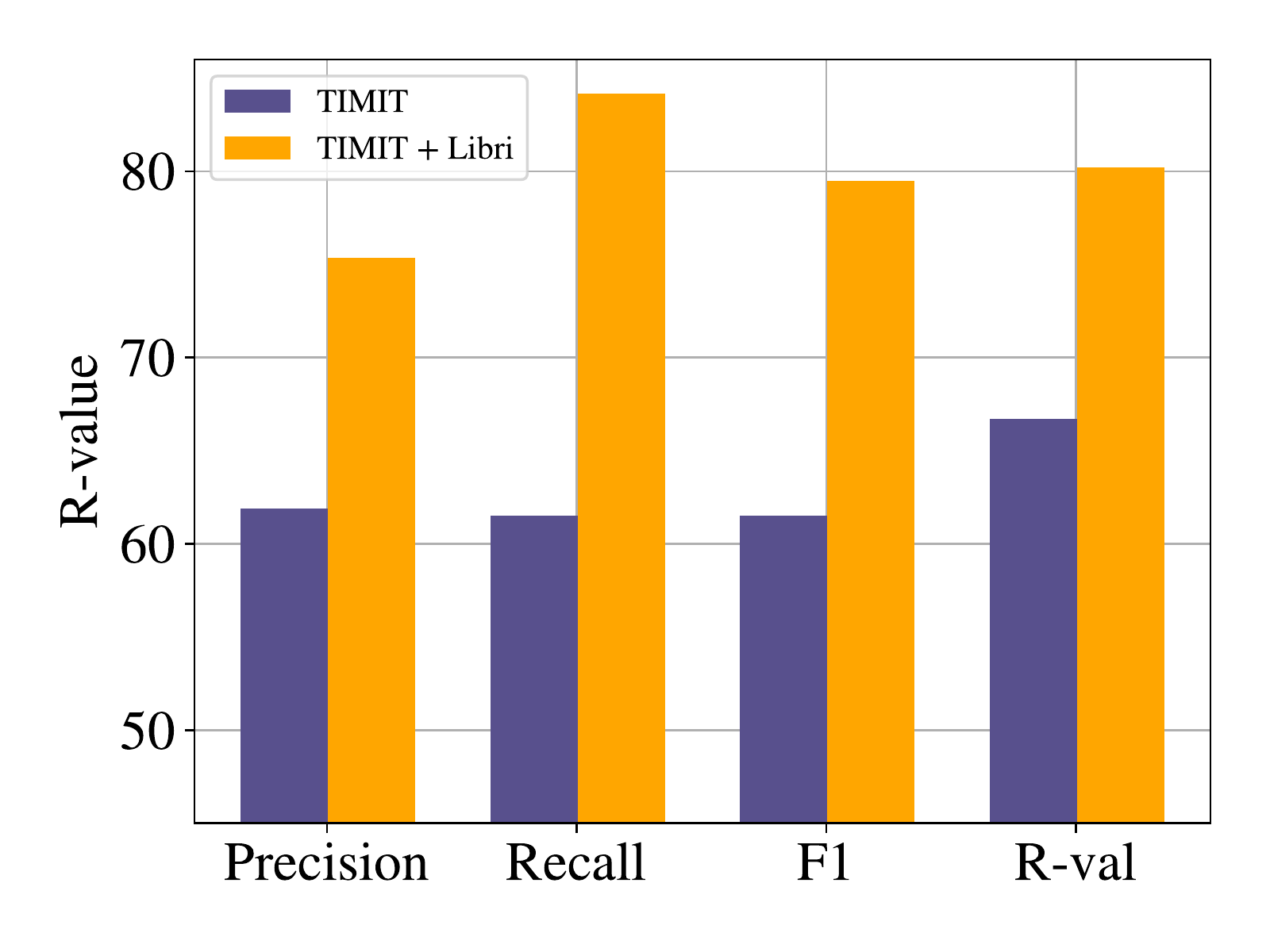}
         \caption{Hebrew}
     \end{subfigure}
     \hfill
     \begin{subfigure}[b]{0.24\linewidth}
         \centering
         \includegraphics[width=\columnwidth]{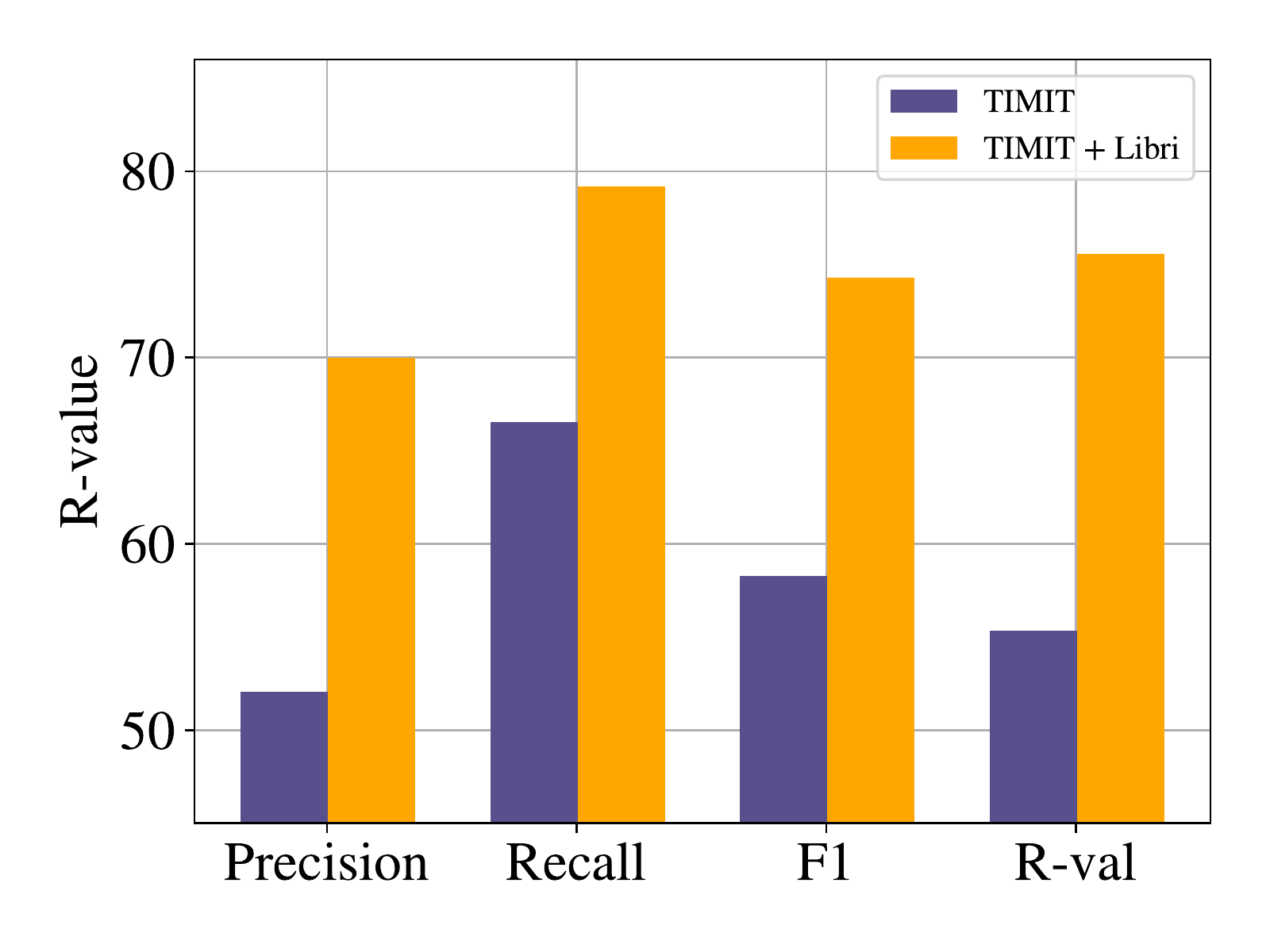}
         \caption{German}
     \end{subfigure}
     \centering
     \begin{subfigure}[b]{0.24\linewidth}
         \centering
         \includegraphics[width=\columnwidth]{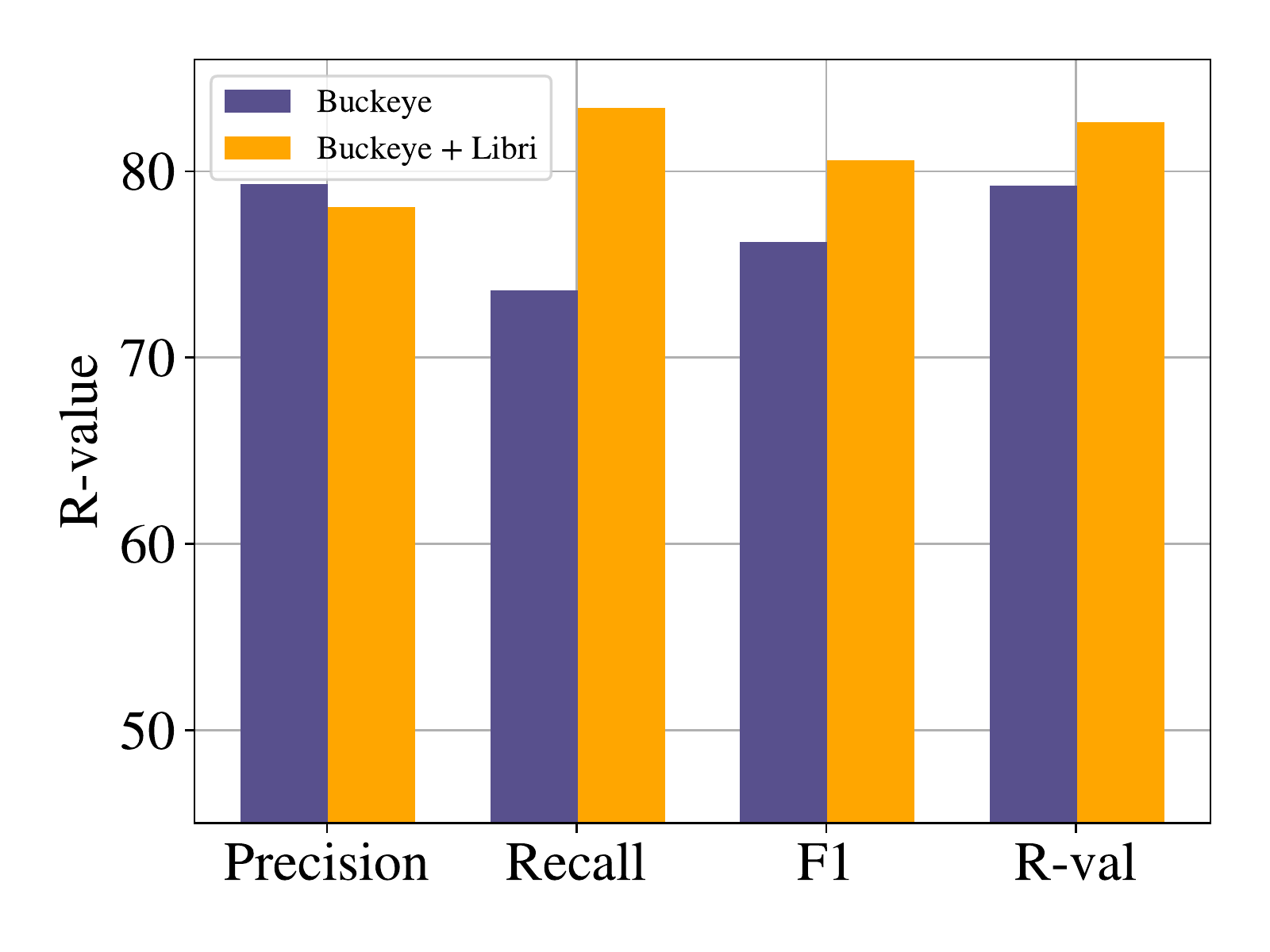}
         \caption{Hebrew}
     \end{subfigure}
     \hfill
     \begin{subfigure}[b]{0.24\linewidth}
         \centering
         \includegraphics[width=\columnwidth]{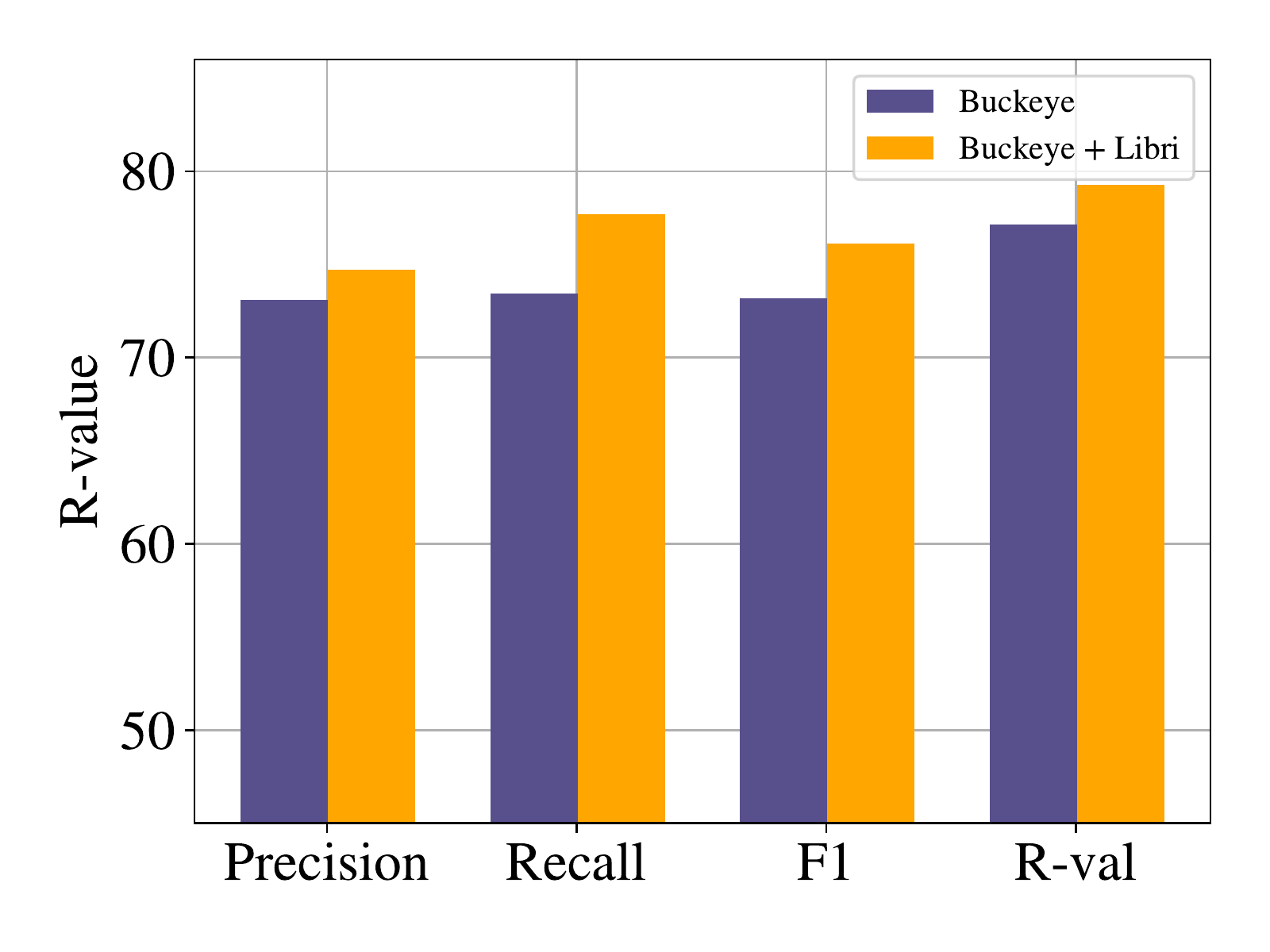}
         \caption{German}
     \end{subfigure}
     \caption{Precision, Recall, F1, and R-value as a function of data added from Librispeech. All models were trained on English training data (sub figures (a) and (b) on TIMIT while sub figures (c) and (d) on Buckeye) and evaluated on both Hebrew and German data sets.}
    \label{fig:r_func_of_libri}
    \vspace{-0.18cm}
\end{figure*}

\subsection{The effect of more training data}
By not relying on manual annotations, SSL methods allow leveraging large unlabeled corpora for additional training data.
In this sub-section we explored the effect of expanding the training set with additional examples from the Librispeech corpus~\cite{panayotov2015librispeech}. We evaluated the model under the following schemes:
(i) training distribution and test distribution match;
(ii) test distribution is different from the training set distribution, but both are from the same language; and
(iii) test and training distributions are from different languages.
In the following experiments, we denote by TIMIT+ and Buckeye+ the augmented versions of TIMIT and Buckeye, respectively. To better match recording conditions we chose different partition from Librispeech to augment TIMIT and Buckeye. For TIMIT+ we used the ``train-clean-100'' partition from Librispeech, while for Buckeye+ we used the ``train-other-500'' partition from Librispeech. 

\paragraph*{In-domain test set}
Results are summarized in Table~\ref{tab:data_ablation}. Surprisingly, the models trained on the augmented training sets showed minor improvements over the original models trained on the TIMIT and Buckeye data sets. In order to better understand the effect of more training data on model performance, we explore the use of out-of-domain test sets in the following paragraphs.

\begin{table}[t!]
	\small
	\centering
	\caption{
	Analysis of model performance on the TIMIT and Buckeye test sets before and after augmenting them with examples from Librispeech.
	}
  \vspace{5pt}
  \label{tab:data_ablation}
  \resizebox{\linewidth}{!}{
  \begin{tabular}{l|l|cccc}
	\toprule
	Training set & Test set & P & R & F1 & R-val    \\
    \midrule
    TIMIT & TIMIT & 83.89 &	83.55 &	83.71 &	86.02   \\
    TIMIT+  & TIMIT & \textbf{84.11} &	\textbf{84.17} &	\textbf{84.13} &	\textbf{86.40}\\
    \midrule
    Buckeye & Buckeye & \textbf{75.78} & 76.86 & 76.31 & 79.69   \\
    Buckeye+  & Buckeye & 74.92 & \textbf{79.41} & \textbf{77.09} &	\textbf{79.82} \\
    \bottomrule
  \end{tabular}
  }
\end{table}
\paragraph*{Out-of-domain test set}
We repeated the experiment from the previous paragraph, however this time with a cross dataset evaluation. In other words, we optimized a model on TIMIT and tested it on Buckeye and vice-versa. Results are summarized in Table~\ref{tab:cross_data}.
\begin{table}[t!]
	\small
	\centering
	\caption{
	Analysis of approach when evaluating the model on a test set that originates from a different distribution than that of the training set.
	}
  \vspace{5pt}
  \label{tab:cross_data}
  \begin{tabular}{l|l|cccc}
	\toprule
	Training set & Test set & P & R & F1 & R-val    \\
    \midrule
    TIMIT & Buckeye & 67.48 &	73.71 &	70.41 &	73.10   \\
    TIMIT+  & Buckeye & \textbf{71.17} & \textbf{81.66} & \textbf{76.05} & \textbf{76.53} \\
    \midrule
    Buckeye & TIMIT & \textbf{86.26} & 79.63 & 82.80 & 84.61  \\
    Buckeye+ & TIMIT & 86.19 &	\textbf{80.10} & \textbf{83.03} & \textbf{84.90} \\
    \bottomrule
  \end{tabular}
  \vspace{-0.2cm}
\end{table}
It can be seen that in cases where the training set and the test set originate from the same distribution (Table~\ref{tab:data_ablation}), adding more data leads to minor improvements in model performance. However, when these are coming from mismatched distributions as seen in Table \ref{tab:cross_data}, adding more data leads to an improvement in performance. For the TIMIT data set, the R-value for the model trained on TIMIT+ was improved by 3.43 points. For the Buckeye data set, we observed a smaller increase in performance. 

\vspace{-0.2cm}
\paragraph*{Multi-lingual evaluation}
Finally, we analyzed the effect of more training data in the multi-lingual setup. To that end, we evaluated the proposed models, trained on TIMIT, TIMIT+, Buckeye, and Buckeye+ (English data), using two data sets from unseen languages. Specifically, we used a Hebrew data set~\cite{benautomatic} and the PHONDAT German data set~\cite{tillmann1993theoretical} as test sets. Figure~\ref{fig:r_func_of_libri} presents the Precision, Recall, F1, and R-value for both data sets with and without additional training data from Librispeech.

Results suggest that utilizing additional unlabeled data yields an increase in performance on unseen languages. For example, when evaluated on the German data set PHONDAT, the TIMIT+ model improved from an R-value of 55.34 to an R-value of 75.58, while on the Hebrew data set the Buckeye+ model improved from an R-value of 79.25 to an R-value of 82.63. Notice, the improvement using TIMIT+ is larger by one order of magnitude comparing to the Buckeye+ improvement. One possible explanation for that is TIMIT being significantly smaller comparing to Buckeye, hence benefiting more from additional data.
These results highlight the importance of additional diverse data sets in cases where there is a mismatch between training set and test set languages. Moreover, this suggests that the representations obtained by the suggested model are not tightly coupled with language-specific features.

\vspace{-0.1cm}
\section{Discussion and future work}
\vspace{-0.1cm}
\label{sec:dis}
In this work we empirically demonstrated the efficiency of self-supervised methods in terms of model performance for the task of unsupervised phoneme boundary detection. Our model reached SOTA results on both TIMIT and Buckeye data sets under the unsupervised setting, as well as showed promising results in terms of closing the gap between unsupervised and supervised methods. Moreover, we empirically demonstrated that using diverse datasets and leveraging more training data produced models with better overall performance on out-of-domain data coming from Hebrew and German.

For future work, we will explore the semi-supervised setting, where we are provided with a limited amount of manually annotated data. Additionally, we will explore the use of the proposed method on low-resource languages and under ``in-the-wild'' conditions. Lastly, we would like to explore the viability of such unsupervised segmentation methods in an unsupervised ASR pipeline.

\bibliographystyle{IEEEtran}
\bibliography{mybib}
\end{document}